# GRADIENTS OF BRAIN ORGANIZATION: SMOOTH SAILING FROM METHODS DEVELOPMENT TO USER COMMUNITY

Jessica Royer[1], Casey Paquola[2], Sofie L. Valk[3], Matthias Kirschner[4], Seok-Jun Hong[5-7], Bo-yong Park[5,8,9], Richard A.I. Bethlehem[10], Robert Leech[11], B. T. Thomas Yeo[12-16], Elizabeth Jefferies[17], Jonathan Smallwood[18], Daniel Margulies[19], Boris C. Bernhardt[1]

1. Montreal Neurological Institute and Hospital, McGill University, Montreal, QC, Canada
2. Institute for Neuroscience and Medicine (INM-7), Forschungszentrum Jülich, Jülich, Germany
3. Max Planck Institute for Human Cognitive and Brain Sciences, Leipzig, Germany
4. Division of Adult Psychiatry, Department of Psychiatry, University Hospitals of Geneva, Thonex, Switzerland
5. Center for Neuroscience Imaging Research, Institute for Basic Science, Suwon, South Korea
6. Center for the Developing Brain, Child Mind Institute, New York, United States
7. Department of Biomedical Engineering, Sungkyunkwan University, Suwon, South Korea
8. Department of Data Science, Inha University, Incheon, South Korea
9. Department of Statistics and Data Science, Inha University, Incheon, South Korea
10. Department of Psychology, University of Cambridge, Cambridge, UK
11. Centre for Neuroimaging Science, King's College London, London, UK
12. Centre for Sleep & Cognition & Centre for Translational Magnetic Resonance Research, Yong Loo Lin School of Medicine, National University of Singapore
13. Department of Electrical and Computer Engineering, National University of Singapore, Singapore
14. Institute for Health & Institute for Digital Medicine, National University of Singapore, Singapore
15. Integrative Sciences and Engineering Programme (ISEP), National University of Singapore, Singapore
16. Martinos Center for Biomedical Imaging, Massachusetts General Hospital, Charlestown, MA, USA
17. Department of Psychology, University of York, York, UK
18. Department of Psychology, Queens University, Kingston, ON, Canada
19. Integrative Neuroscience and Cognition Center (UMR 8002), Centre National de la Recherche Scientifique (CNRS) and Université de Paris, Paris, France

Address correspondence to:

Jessica Royer, PsyD

jessica.royer@mail.mcgill.ca

**ABSTRACT**

Multimodal neuroimaging grants a powerful *in vivo* window into the structure and function of the human brain. Recent methodological and conceptual advances have enabled investigations of the interplay between large-scale spatial trends – or *gradients* – in brain structure and function, offering a framework to unify principles of brain organization across multiple scales. Strong community enthusiasm for these techniques has been instrumental in their widespread adoption and implementation to answer key questions in neuroscience. Following a brief review of current literature on this framework, this perspective paper will highlight how pragmatic steps aiming to make gradient methods more accessible to the community propelled these techniques to the forefront of neuroscientific inquiry. More specifically, we will emphasize how interest for gradient methods was catalyzed by data sharing, open-source software development, as well as the organization of dedicated workshops led by a diverse team of early career researchers. To this end, we argue that the growing excitement for brain gradients is the result of coordinated and consistent efforts to build an inclusive community and can serve as a case in point for future innovations and conceptual advances in neuroinformatics. We close this perspective paper by discussing challenges for the continuous refinement of neuroscientific theory, methodological innovation, and real-world translation to maintain our collective progress towards integrated models of brain organization.

## INTRODUCTION

Foundational neuroanatomical studies performed over the course of the 20<sup>th</sup> century identified graded changes in cellular organisation across the cortex (Abbie, 1940; Bailey & von Bonin, 1951; Dart, 1934; Sanides, 1962). These *"gradients"* variably capture step-wise changes across several cortical areas, such as decreasing laminar differentiation along the cortical visual hierarchy (Hilgetag & Grant, 2010), as well as smooth transitions between areas, notably reflected in the decreasing size of pyramidal neurons moving anterior from the primary motor cortex (Bailey & von Bonin, 1951). Moreover, some gradients are repeated in different brain lobes (García-Cabezas et al., 2019; Margulies et al., 2016), revealing large-scale *"axes"* of brain organization, that is the organization of multiple – and often distributed – cortical areas along a continuous feature representation (Huntenburg et al., 2018).

This perspective has recently gained traction in neuroscience, as researchers using machine learning to reduce the dimensionality of brain features have found that resulting axes often resemble spatially-constrained gradients. One widely used approach involves the computation of an affinity matrix to quantify inter-regional similarity in one or several features, followed by dimensionality reduction to generate a continuous ordering of matrix nodes in a lower dimensional manifold space (Bajada et al., 2020; Haak et al., 2018; Vos de Wael et al., 2020). These methods have been applied to numerous data modalities in human and non-human species, particularly using magnetic resonance imaging (MRI), to uncover gradients in measures of brain microarchitecture, anatomy, and function. For instance, previous work has shown gradients in cortical cyto- and myeloarchitecture (Burt et al., 2018; Paquola, Benkarim, et al., 2020; Paquola, Seidlitz, et al., 2020; Paquola et al., 2019; Royer et al., 2020; Saberi et al., 2023), gene expression (Dear et al., 2022; Froudist-Walsh et al., 2023; Vogel et al., 2020), receptor architecture (Hansen et al., 2022; Luppi et al., 2023), cortical thickness (Hettwer et al., 2022; Valk et al., 2020; Wagstyl et al., 2015), diffusion tractography-based structural connectivity (Blazquez Freches et al., 2020; Park, Vos de Wael, et al., 2021; Vos de Wael et al., 2021), and intrinsic functional connectivity computed from resting-state and task-based functional MRI signals (Cabalo et al., 2023; Ito et al., 2020; Margulies et al., 2016; Vos de Wael et al., 2018). By depicting these properties along a low-dimensional and continuous coordinate system, gradients have provided a valuable framework to integrate distinct neural motifs (Mars et al., 2021). Indeed, findings from these techniques frequently reveal confluent spatial trends across modalities, indicative of their ability to capture core principles of brain organization. Expanding this idea, quantifying reorganizations of this coordinate system has granted novel insights into changes occurring as a function of typical and atypical development and aging (Benkarim et al., 2021; Bethlehem et al., 2020; Hong et al., 2019; Larivière et al., 2020; Park, Bethlehem, et al., 2021; Petersen et al., 2022; Setton et al., 2023), neurological and psychiatric disease (Cabalo et al., 2023; Hettwer et al., 2022; Meng et al., 2021; Royer et al., 2023; Wan et al., 2023), cognitive processes (Gao et al., 2022; McKeown et al., 2020; Valk et al., 2023), and the evolution of our species (Fulcher et al., 2019; Xu et al., 2020).

This rich literature highlights how gradient methods naturally lend themselves to numerous data modalities and populations. It also shows how they can help answer questions about fundamental neuroscientific principles and may be translated to clinical contexts. Such flexibility invites multiple perspectives from an inherently multidisciplinary user community. Nonetheless, the adoption of these methods could not have been possible without the implementation of tailored strategies to facilitate the initiation of users with diverse backgrounds, skillsets, and unequal access to data and computational resources. This perspective paper highlights and discusses pragmatic and human elements facilitating the widespread visibility, application, and development of brain gradients in neuroscience. We particularly discuss the crucial role played by deeply rooted open neuroscience practices, and the critical role that early-career researchers have played in promoting these tools through various channels (**FIGURE 1**). Collectively, this review underlines how the adoption of gradient methods by the community reflects the increasing synergy between strong domain expertise in neuroscience and advanced computational skillsets to construct unified, integrative models of multiscale brain organization.

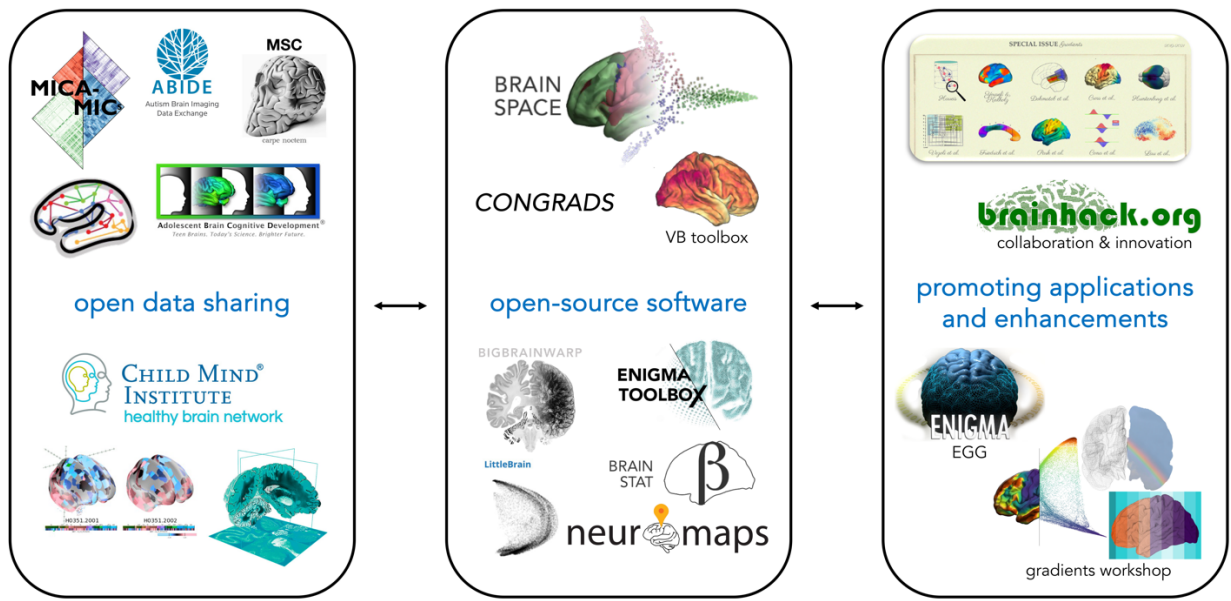

**Figure 1. Resources enabling the development of new methods in neuroinformatics**. Advances in the analysis of brain gradients have been catalyzed by open neuroscience practices, particularly data sharing and collaborative software development, as well as initiatives promoting gradient methods and their applications. We argue that the growing adoption of these techniques can serve as a case study for future methodological and conceptual advances in neuroinformatics: By creating a self-sustaining user community, these elements helped establish gradient methods as a key approach to interrogate and integrate large multimodal neuroimaging datasets and map multiscale brain organization.

**OPENING MULTIMODAL AND MULTISCALE STUDIES OF BRAIN ORGANIZATION**

In recent years, neuroscience has seen mounting availability of open datasets providing several imaging modalities to probe human brain structure and function *in vivo* (Alexander et al., 2017; Casey et al., 2018; Di Martino et al., 2017; Di Martino et al., 2014; Gordon et al., 2017; Miller et al., 2016; Poldrack et al., 2015; Royer et al., 2022; Thompson et al., 2014; Van Essen et al., 2013). In parallel, datasets leveraging histological staining and transcriptomics from *post mortem* samples co-registered to standard spaces commonly used in neuroimaging have enabled novel explorations of multiscale human brain organization (Alkemade et al., 2022; Amunts et al., 2013; Hawrylycz et al., 2012). The aggregation of such diverse, openly available datasets has generated new opportunities for unified and integrative models of brain architecture, accelerating both theoretical and technical advances in neuroimaging and neuroinformatics (Milham, Craddock, et al., 2018).

Gradient techniques constitute a powerful approach to bridge different neurobiological features and scales, and thus provide conceptual and methodological means to harness growing data complexity in neuroscience (Huntenburg et al., 2018). This was demonstrated in several recent studies leveraging low-dimensional depictions of neural organization to explore systematic co-variations in structural and functional brain properties expressed at distinct scales, notably in investigations of the structural underpinnings of transmodal functional systems (Margulies et al., 2016; Paquola et al., 2019; Xu et al., 2020). For instance, initial applications of gradient methods to the openly available Human Connectome Project dataset (Van Essen et al., 2013) uncovered a principal gradient of intrinsic functional connectivity differentiating sensorimotor systems from transmodal networks encompassing default-mode and paralimbic structures, recapitulating foundational models of the cortical hierarchy initially formulated in non-human primates (Margulies et al., 2016; Mesulam, 1998). This hierarchy is strongly reflected in neocortical geometry, as the topography of this functional gradient reflected measures of geodesic distance between sensory and transmodal subregions (Leech et al., 2023; Margulies et al., 2016; Smallwood et al., 2021; Wang, Krieger-Redwood, Lyu, et al., 2023; Wang, Krieger-Redwood, Zhang, et al., 2023). Furthermore, the spatial patterning of this gradient is in line with a proposed macroscale mechanism allowing transmodal networks to support cognitive functions decoupled from our immediate external environment, or the "here and now" (Poerio et al., 2017; Smallwood & Schooler, 2015; Smallwood et al., 2013).

These techniques have also informed our understanding of the evolution of neural function in transmodal cortices, via the joint embedding of cross-species functional connectivity data from two openly available datasets (Milham, Ai, et al., 2018; Van Essen et al., 2013). Here, extracting the most similar functional dimensions in both humans and macaques enabled comparisons of regional functional profiles across both species within a homologous space. These findings could thus shed light on aspects of human cognition supported by spatially distributed association cortices and their divergence from close evolutionary ancestors (Xu et al., 2020).

Relatedly, gradients constructed from intrinsic functional connectivity data and myelin-sensitive MRI contrasts highlighted a progressive dissociation of functional and microstructural hierarchies in the human brain peaking in transmodal default mode and fronto-parietal networks (Paquola et al., 2019). Notably, the microstructural gradient topography uncovered *in vivo* was replicated in histological data provided by *BigBrain*, an ultra-high-resolution volumetric reconstruction and cortical surface segmentation of a sliced and stained *post mortem* human brain (Amunts et al., 2013; Paquola et al., 2019). This work showed that spatial locations of strong structure-function divergence co-localized with key aspects of human cognition such as cognitive control and social cognition, supporting the idea that transmodal networks may be relatively decoupled from hierarchical constraints, enabling these regions to assume more flexible functional roles (Paquola et al., 2022; Paquola et al., 2019; Valk et al., 2022).

These techniques have also provided insights into the macroscale organization of specific cortical structures, revealing patterns of structure-function dissociation in regions with particularly complex architectures. For instance, recent work has shown only partial overlap between microstructural and functional gradients of the insula, a hub for somatosensory, socio-affective, and cognitive processes in humans (Royer et al., 2020; Tian & Zalesky, 2018). In some structures, these principal axes may even run orthogonally. While transcriptomic and functional connectivity gradients are seemingly expressed in an anterior-posterior fashion in the hippocampus (Vogel et al., 2020; Vos de Wael et al., 2018), prominent cytoarchitectural variations are rather seen in the proximal-distal direction, following the topography of its subfields (Genon et al., 2021; Paquola, Benkarim, et al., 2020).

These examples highlight how gradient approaches can offer a formal framework for multimodal and multiscale neuroscience. Although the growing adoption of these techniques can be understood in the context of emerging large-scale open neuroscience initiatives, we also emphasize that their application has directly informed our understanding of brain organization. As the development of these methods has so strongly benefitted from data sharing, we encourage the user community to continue injecting their solutions back into open neuroscience ecosystems. The contribution of open-source software for the computation of brain gradients has been instrumental in this regard (see next section). Nonetheless, open and centralized gradient map repositories associated with common neuroimaging datasets remain to be implemented. Such resources will become increasingly important with growing sample sizes and higher data resolution, underscoring the need for more efficient management of computational efforts (Horien et al., 2021). These initiatives may be implemented in different forms: Although group-average gradient maps are already available in open software tools for select datasets and modalities (Guell et al., 2019; Larivière et al., 2023; Larivière et al., 2021; Markello et al., 2022; Paquola et al., 2021), open repositories may also include individual-level maps to enable more flexible integration within existing analytical pipelines (Royer et al., 2022). Such post-processed data sharing platforms are still in their infancy and constitute interesting opportunities for gradient methods development.

**OPEN-SOURCE SOFTWARE TO UNCOVER BRAIN GRADIENTS**

Large-scale data sharing efforts must be paralleled by methodological advances to process, aggregate, and contextualize associated features. As datasets increase in complexity and size, the tools required to explore the rich information they contain must scale accordingly, and quickly highlight a need for shared, collaborative solutions (Freeman, 2015; Harding et al., 2023). The availability of open software to generate and analyze brain gradients is no exception to this view. The development of such tools, their thorough documentation, and the openness of developers to contributions and enhancements from the user community have been instrumental in facilitating the accessibility of gradient methods to researchers across a wide range of subdisciplines in neuroscience. These open practices have enabled gradient techniques to move beyond the borders of single research groups, fostering openness to methodological innovation and new application perspectives from various subdisciplines in brain imaging research. Crucially, these initiatives leverage established software development practices with clear potential to accelerate progress in neuroscience (Gewaltig & Cannon, 2014). The current section reviews and discusses this progress, with a focus on (1) available software packages designed for the computation of brain gradients and (2) various toolboxes including pre-computed gradients maps to facilitate the contextualization of neuroimaging findings.

One widely used toolbox for the computation of brain gradients is *BrainSpace* (https://github.com/MICA-MNI/BrainSpace). Available in both MATLAB and python, this open-access toolbox allows for the identification and analysis of gradients from any bioimaging feature that can be represented in matrix form (Vos de Wael et al., 2020). Since its publication, BrainSpace has been applied by several research groups to study transcriptomic (Dear et al., 2022), metabolic (Luppi et al., 2023), structural (Paquola, Seidlitz, et al., 2020; Park, Vos de Wael, et al., 2021; Saberi et al., 2023; Valk et al., 2020), and functional (Huang et al., 2023; Katsumi et al., 2022;

Sporns et al., 2021; Timmermann et al., 2023) brain properties, as well as interactions between these modalities (Tong et al., 2022; Valk et al., 2022) and the statistical significance of relationships between spatially smooth feature maps (Coletta et al., 2020; Raut et al., 2021), making it a widely used analytical resource in brain imaging. This package offers several options for customized gradient analysis, notably allowing for different kernels, dimensionality reduction techniques, alignment methods, as well as null models to examine statistical significance in the spatial correlation between brain maps. Alongside detailed documentation and practical tutorials (https://brainspace.readthedocs.io/), these features offer a highly flexible and approachable analysis environment within BrainSpace, facilitating the initiation of new users and continuous software development and maintenance.

Next, *CONGRADS* (https://github.com/koenhaak/congrads) provides a python-based infrastructure for the investigation of gradients of functional connectivity data within user-defined region-of-interests (Haak et al., 2018). Indeed, the authors motivate their work by highlighting the widespread presence of connection topographies (or *connectopies*) in the human brain, by which neighbouring areas within a given structure connect with nearby locations in distant brain structures. This toolbox follows this principle via the implementation of Laplacian eigenmaps, a non-linear manifold learning technique, to recover local connectopies and facilitate investigations of individualized transitions in regional connectivity patterns. By cross-referencing functional connectopies mapped in primary visual and motor cortices with corresponding regional retinotopic and somatotopic organization, the authors could furthermore provide evidence for the biological validity of this technique. Previous work has harnessed this approach for its potential to recover overlapping principles of subregional connectome organization (Haak & Beckmann, 2020; Mars et al., 2018), notably in the subcortex (Marquand et al., 2017; Tian et al., 2020), and within cortical structures such as the insula (Tian & Zalesky, 2018; Tian et al., 2019) and hippocampus (Przeździk et al., 2019), testament to its versatility.

The *Vogt-Bailey toolbox* (VB toolbox; https://github.com/VBIndex), also available in both MATLAB and python, implements similar gradient analysis pipelines to CONGRADS and BrainSpace. However, a key contribution of this toolbox consists in it providing quantitative means to measure the sharpness of a given brain area defined from selected imaging features (Bajada et al., 2020; Ciantar et al., 2022). This contribution builds on longstanding microarchitectural perspectives on cortical organization, balancing the views of the Vogt school, which emphasized the existence of sharp boundaries between cortical regions (Brodmann, 1909; Vogt & Vogt, 1903), and those of Bailey and Von Bonin arguing for a rather graded cortex (Bailey & von Bonin, 1951). The resulting *VB index* reflects the "gradedness" of the input feature across multiple scales, namely for the entire cortex, unique clusters of regions of interest, or at a vertex-level. Building on foundational perspectives in systems and network neuroscience, the metric introduced in this software thus introduces an innovative approach to quantitatively assess spatial properties of brain gradients.

The growing accessibility of gradient methods and findings has also been facilitated by the inclusion of pre-computed gradient maps in several recently developed software packages. Gradients computed using established, openly available datasets are offered as part of these toolboxes to ground new imaging findings within topographic principles of human brain organization. This approach was implemented in *neuromaps* (https://github.com/netneurolab/neuromaps), a toolbox and brain map repository for accessing, transforming and decoding structural and functional brain data (Markello et al., 2022). Neuromaps streamlines transformations between several standard surface- and volumetric template spaces while providing an open and collaborative platform for user contributions of new post-processed data. As a result, neuromaps provides an impressive arsenal of tools for the development of integrative views of brain architecture. The neuromaps paradigm was recently extended to the hippocampal region in the form of the hippomaps repository (https://hippomaps.readthedocs.io/), which aggregates histological as well as imaging derived maps in a common hippocampal coordinate system. Another such unified transformation-contextualization environment is proposed in *BigBrainWarp* (https://github.com/caseypaquola/BigBrainWarp), a unique toolbox for the integration of *post mortem* histological staining data (Amunts et al., 2013) with neuroimaging and other neurobiological modalities (Paquola et al., 2021). Additional contextualization-focused software, including the *ENIGMA toolbox* (https://github.com/MICA-MNI/ENIGMA) and *BrainStat* (https://github.com/MICA-MNI/BrainStat), both propose mapping and plotting tools to stratify input brain features according to the topography of gradient maps (Larivière et al., 2023; Larivière et al., 2021). Similarly, *LittleBrain* (https://github.com/xaviergp/littlebrain) provides tools to interpret cerebellar neuroimaging findings along its functional connectivity gradients (Guell et al., 2019; Guell et al., 2018).

This recent explosion of available tools to compute and visualize brain gradients underlying a user's choice of data modality, as well as the ever-growing application of gradient maps to contextualize new findings in neuroimaging emphasize the central role that these methods now occupy in contemporary neuroscience research. However, new tools are only useful if they are actively put to work and rigorously tested by their pool of potential users, creating an engaged community to apply, promote, and improve the methods they implement. This requires coordinated efforts in the direction of accessibility and continuous support to create a truly inclusive community, where new users are encouraged to integrate these tools within existing analytical pipelines and conceptualize new avenues of inquiry that can be addressed with gradients methods.

**PROMOTING DEVELOPMENT, APPLICATIONS, AND NEW PERSPECTIVES**

Gradients have emerged from and contributed to neuroscientific research by leveraging existing data resources and giving back software and processed feature maps to the community. Beyond these more technical contributions, we also underline the importance of selecting appropriate outlets to promote these tools and the innovations they generate. These knowledge transfer methods can be broadly organized according to traditional vs. disruptive disseminations strategies, which we review in the present section. Importantly, all efforts were put forward in the spirit of inclusivity and openness, highlighting the human aspects of implementing and promoting new methods in neuroinformatics.

Important contributions to the gradient literature were collected in a recent special issue on the topic (Bernhardt et al., 2022). This special issue reflected the rapidly expanding literature on brain gradients, with 36 unique contributions covering methodological advances in gradient mapping (Bajada et al., 2020; Burt et al., 2020; Dadi et al., 2020; Dohmatob et al., 2021; Glomb et al., 2020; Hong et al., 2020; Nenning et al., 2020; Patel et al., 2020), conceptual perspectives on these techniques (Haak & Beckmann, 2020; Haueis, 2021), as well as the application of gradients to explore neuroanatomical organization (Lefco et al., 2020; Müller et al., 2020; Saadon-Grosman et al., 2020), structure-function coupling (Cona et al., 2021; Friedrich et al., 2020; Masouleh et al., 2020; Royer et al., 2020; Vezoli et al., 2021; Waymel et al., 2020; Yang et al., 2020), dynamics and states (Cross et al., 2021; Liu et al., 2020; Mitra et al., 2020; Park, Vos de Wael, et al., 2021; Yousefi & Keilholz, 2021), cognitive systems (Ito et al., 2020; Lanzoni et al., 2020; McKeown et al., 2020; Viviani et al., 2020; Wang et al., 2020), evolution (Huntenburg et al., 2021; Lau et al., 2021; Xu et al., 2020), and maturation and aging (Ball et al., 2020; Bethlehem et al., 2020). By collecting current innovations in the field, this special issue helped cement the position of brain gradients as an established perspective in network neuroscience while highlighting open questions and challenges for the field.

In terms of conference activity, the annual Gradients of Brain Organization workshop (also known as the "gradients workshop") is recognized as a part of a series of events leading up to the annual meeting of the Organization for Human Brain Mapping (OHBM). As a pre-OHBM event, the gradients workshop has gained strong visibility within the community, gathering 12-15 speakers for full length talks, 30-40 submitted abstracts for the poster session each year, and up to 300 attendees (combining virtual and in-person participants). As the workshop approaches its fifth edition in 2024, the organizing committee remains committed to equity, diversity, and inclusion in their efforts to constitute the workshop's yearly speaker line-up. The committee indeed implements strict rules on gender and geographic diversity of invited speakers. Furthermore, content is renewed with fresh perspectives every year, as organizers avoid repeated invitations of the same speakers. In addition to elevating a diversity of voices involved in the gradient community, this policy has fostered the growth of a self-sustaining community where early-career researchers are encouraged to share their vision for the field. The gradient workshop thus remains, year after year, a forum for researchers at different career stages representing a range of neuroscientific subdisciplines to discuss methods development, new applications, and findings relating to brain gradients.

Gradient methods were also brought to the larger neuroimaging community as part of different Brainhack projects (https://brainhack.org/). Brainhack is an established, alternative meeting format emphasizing inclusivity, collaboration, and education in a project-oriented manner (Gau et al., 2021). Several published projects in the gradients literature were conceptualized or initiated as part of Brainhack. These initiatives notably led the application of gradients to clinical populations such as autism spectrum disorder, offering evidence of macroscale, system-level imbalance in functional circuits in this condition (Bethlehem et al., 2017; Hong et al., 2019). More recent collaborative models harnessing brain gradients have emerged via the Enhancing NeuroImaging Genetics through Meta-Analysis (ENIGMA) Consortium, specifically the ENIGMA Gradient Group (EGG: https://enigma.ini.usc.edu/ongoing/enigma-gradient/). While leveraging existing datasets, EGG also aims to initiate data acquisition frameworks for microstructure-sensitive and functional imaging data, in addition to developing and documenting open tools for the analysis and contextualization of brain gradients.

**THE FUTURE OF A SELF-SUSTAINING COMMUNITY**

As datasets in neuroscience and neuroimaging increase in size and complexity, new tools for the multimodal and multiscale integration of these data have progressively been deployed. As a result, researchers wishing to harness these resources must now demonstrate expertise in specific areas of neuroscientific inquiry as well as the necessary computational skills to interrogate emerging datasets. Brain gradients offer a unique case study for this transition. This perspective paper specifically outlines the elements leading to widespread recognition and adoption of brain gradients by the neuroscience community. Established open science practices in neuroimaging and network neuroscience have played a key role in this process by granting more equitable access to data and software across the field. In addition, the promotion of gradient methods and their applications via more traditional as well as more innovative learning and knowledge transfer formats have created an inclusive and diverse community of users and contributors. Building on this foundation, brain gradients have granted new opportunities to study fundamental principles of brain organization bridging structure, function, disease, development, and evolution within a self-sustaining and ever-growing user base.

Several challenges remain, however, in sustaining this gradient user community. From a technical standpoint, important consideration should be directed towards the sustainability of software tools supporting the progression of our field. As is common in academic software development contexts, sustainable continuation plans are often poorly defined or lacking, and tools become quickly deprecated once trainees leading a given project move on to other positions (Gewaltig & Cannon, 2014). Notably, despite the relative infancy of tools to generate brain gradients, some software reviewed in the present paper has seen no repository activity, including commits and responses to issues, in several years. As our community relies heavily on a common set of open software for the analysis of brain gradients, concrete plans for long-term sustainability will have to be put in place. This continued support will enable the development of flexible approaches to integrate gradient methods with emerging brain mapping techniques.

Even with this growing number applications, the field still lacks systematic and established practices for gradient parameterization, namely regarding hyperparameter tuning and technical choices in the gradient computation pipeline (*e.g.,* choice of kernel and dimentionality reduction technique). Such efforts are only beginning to be deployed: Although previous investigations remain specific to single data modalities (e.g., functional connectivity from rs-fMRI), they have demonstrated ability to reliably predict behavioural patterns, and can do so more accurately than models relying on connectivity weights (Hong et al., 2020). Nonetheless, systematic, quantitative approaches to inform parameters tuning and selection should be developed and made openly available to the community. This promises to facilitate robust and reproducible methodological frameworks, and their systematic implementation across research groups, data modalities, and populations. As gradient methods are increasing deployed to investigations of neuropsychiatric disease, encouraging more systematic parameter benchmarking will be necessary to assess the potential of brain gradients to serve as reliable biomarkers of patient phenotypes.

Future applications of these methods may support the conceptualization of biologically-inspired machine learning and artificial intelligence-based models of behaviour and brain organization. At the most basic level, brain gradients may streamline feature selection in machine learning models by reducing the high dimensionality of brain graphs (Hong et al., 2020; Shevchenko et al., 2024). However, gradient methods may also be directly integrated in certain model architectures or assist in conceptual interpretation of their findings. For instance, in the case of artificial intelligence, artificial neural networks (ANNs) are increasingly recognized in their potential to inform investigations and models in systems neuroscience, opening new areas of inquiry in the field (Kanwisher et al., 2023; Richards et al., 2019). This link between artificial and neural systems may take different forms, where brain gradients could notably assist in defining the architecture of ANNs to reflect the hierarchical processing streams encoded in complex brain networks, or help interpret why the topography of certain functional systems is expressed as it is in brain (Kanwisher et al., 2023). These approaches could thus assist in quantifications of the added value of incorporating brain gradients, alongside complementary parcellation-based perspectives, into our investigations of brain organization.

In addition to these challenges, the field is also presented with interesting opportunities: With growing access to open data resources, software tools, and training opportunities to compute and analyze brain gradients, researchers possess all necessary resources to steer the field in new directions of scientific inquiry. Although meaningful metrics to quantify the impact and sustainability of gradient methods and their applications should be clearly defined, the community holds the power to sustain itself for years to come.